\RequirePackage{fix-cm}
\documentclass[smallextended]{svjour3}       % onecolumn (second format)
\smartqed  % flush right qed marks, e.g. at end of proof
\usepackage{graphicx}
\usepackage{natbib}
\usepackage{amsfonts}
\usepackage{amsmath}
\usepackage{hyperref}
\usepackage{bm}
\usepackage[all]{xy}
\newcommand\Tstrut{\rule{0pt}{2.6ex}}  
\newcommand\Bstrut{\rule[-0.9ex]{0pt}{0pt}}
\newcommand{\Gr}{\mathcal{G}}
 
  \newcommand{\E}{\mathbb{E}}
   \newcommand{\V}{\mathbb{V}}
 \newcommand{\dr}{\textnormal{d}}
 \newcommand\independent{\protect\mathpalette{\protect\independenT}{\perp}}
\def\independenT#1#2{\mathrel{\rlap{$#1#2$}\mkern2mu{#1#2}}}
\journalname{Annals of Operations Research}
\begin{document}

\title{Coherent combination of probabilistic outputs for group decision making: an algebraic approach\thanks{We acknowledge that J.Q. Smith was partly supported by EPSRC grant EP/K039628/1 and The Alan Turing Institute under EPSRC grant EP/N510129/1, whilst E. Riccomagno was supported by the GNAMPA-INdAM 2017 project.}
%Grants or other notes
}

\titlerunning{Coherent combination of probabilistic outputs for group decision making}        % if too long for running head

\author{Manuele Leonelli         \and
        Eva Riccomagno \and Jim Q. Smith
}


\institute{Manuele Leonelli \at
              School of Mathematics and Statistics, University of Glasgow, UK \\
              \email{manuele.leonelli@glasgow.ac.uk}           %  \\
           \and
          Eva Riccomagno \at
              Dipartimento di Matematica, Universit\`{a} degli Studi di Genova, Italia\\
              \email{riccomagno@dima.unige.it}
              \and 
              Jim Q. Smith \at
              Department of Statistics, University of Warwick, UK\\
              \email{j.q.smith@warwick.ac.uk}
}

\date{Received: date / Accepted: date}

\maketitle

\begin{abstract}
Current decision support systems address domains that are heterogeneous in nature and becoming progressively larger. Such systems often require the input of expert judgement about a variety of different fields and an intensive computational power to produce the scores necessary to rank the available policies. Recently,  integrating decision support systems have been introduced to enable a formal Bayesian multi-agent decision analysis to be distributed and consequently efficient. In such systems, where different panels of experts oversee disjoint but correlated vectors of variables, each expert group needs to deliver only certain summaries of the variables under their jurisdiction to properly derive an overall score for the available policies.  Here we present an algebraic approach that makes this methodology feasible for a wide range of modelling contexts and that enables us to identify the  summaries needed for such a combination of judgements. We are also able to demonstrate that coherence, in a sense we formalize here, is still guaranteed when panels only share a partial specification of their model with other panel members. We illustrate this algebraic approach by applying it to a specific class of Bayesian networks and demonstrate how we can use it to derive closed form formulae for the computations of the joint moments of variables that determine the score of different policies. 

\end{abstract}

\section{Introduction}
\label{sec:intro}
Although still being refined, probabilistic decision support tools for single agents are now well developed and used in practice in a variety of domains.  One of the most common probabilistic models for multivariate systems are Bayesian networks (BNs) \citep{Jensen2013,Pearl1988} and their dynamic and object-oriented extensions \citep{Koller1997,Murphy2002}.  However,  these are not the only frameworks around which probabilistic models have been built. Other well-established models comprise, among others, Bayesian hierarchical spatio-temporal models \citep{Blangiardo2015}, asymmetric probability trees \citep{Smith2008} and probabilistic emulators \citep{Kennedy2001}.

However, the size and complexity of current applications often require that supporting systems  consist of component modules which, encoding the judgements of panels of domain experts, describe a particular sub-domain of the overall system.  In these contexts decision makers need a tool that can coherently paste together the outputs of each of these modules to provide a comprehensive picture of the whole process. Often in practice, because of both computational and methodological constraints, the modules' outputs end up being collated together in a simple, essentially deterministic way by transferring from one module to another a single vector of means about what might happen and hence effectively ignoring any associated uncertainty. However, such a na\"{i}ve method can be very misleading and guide decision makers to choose a  non-optimal course of action \citep[see e.g.][]{Leonelli2013, Leonelli2015}.
 
 Recently, integrating decision support systems  (IDSSs) \citep{Leonelli2015,Smith2015} have been defined to extend coherence requirements traditionally applied within a Bayesian decision support system for single agents so that these apply to this new multi-expert setting. IDSSs embed a methodology, similar to a standard Bayesian one, where decisions can be guaranteed to be coherent, i.e. expected utility maximising for some utility and probability distribution derived from individual but connected suites of models.  Before briefly reviewing the theory of IDSSs in Section \ref{sec:IDSS}, we discuss in Section \ref{sec:food} a domain of application where we have found it necessary to knit together a suite of models. We then introduce in Section \ref{sec:example} a real-world example that illustrates our methodology. In Section \ref{sec:contributions} we highlight the contributions of the paper.

\section{Motivation and contribution} 
\subsection{UK household food security}
\label{sec:food}
Food security, once thought to be a problem confined to low-income countries, is increasingly being recognised as a matter of concern in wealthy nations like the UK \citep{Dowler2015}. At a country level the UK is among the most food secure in the world, but at household level the number of individual with  limited or uncertain availability of nutritionally adequate and safe foods is increasing rapidly \citep{Loopstra2015}.  At first glance, UK household food security may seem to be a simple case of demand and supply. On closer inspection though the food system is shown to be highly complex, especially from the viewpoint of policymakers, who endeavour to intervene on the system in order to provoke specific ameliorating responses \citep{Dowler2015,Drewnowski2004}.

The food system is global, multifaceted and influenced by a huge number of public and private actions and uncontrolled factors such as weather, pests and disease. This leads to a great deal of uncertainty about the effectiveness of any one policy. One of the authors has been involved, in partnership with Warwickshire County Council, UK, in the development of an IDSS to support decision-making around household-level food poverty.  An overall description of the highly heterogeneous food system requires the judgements from different panels of experts in diverse disciplines including insights about factors elevating the risk to food security of households (from sociologists and local authorities), judgements about the effects of malnutrition on the population (from doctors and nutritionists), estimates of the availability of food in supermarkets and other outlets (from supply chain experts) and forecasts of the yield of crops in a particular season (from crop experts).

 Unless properly structured, this expert information is liable to conflict since two or more panels can sometimes deliver contradicting expert judgements about a shared random variable. If such contradictions are admitted then the system's coherence is obviously threatened and its outputs become compromised. For instance, both estimates of cost of oil and weather forecasts affect food production, food transport and the ability of households to access food. If these latter variables are under the jurisdiction of different panels, any integrating system should surely embed common estimates of distributions over the cost of oil and weather forecasts and not contradicting ones. Otherwise how could it ever be coherent and justifiable?

Whilst numerous systems to model aspects of the composite process exist, such as for supermarket locations and food demand forecasting \citep{Efendigil2009,Hernandez2000}, the complex problem of developing a shared methodology to guide the accommodation of diverse expertise and that provide enough information  to evaluate the efficacy of various policies designed to address food poverty issues has been attempted only recently \citep{Barons2015,Smith2015}.  Protocols to guide this probabilistic integration have been discussed only in the context of BN models \citep{Johnson2012,Mahoney1996}. The food security  and other applications, as for instance nuclear emergency management \citep{Leonelli2013,Leonelli2015}, have motivated the methodological developments we present below.

\subsection{How an IDSS works}
\label{sec:IDSS}
Although the decomposition of a complex system into connected but separated components overseen by different panels of experts may seem reasonable in most cases, various conditions need to be entertained for an IDSS to be justifiable. More specifically it can be argued that an IDSS requires the following to hold:
\begin{itemize}
\item the decision centre responsible for the implementation of any policy needs to consist of individuals who act collaboratively and strive to behave as a single coherent unit would. In the food poverty \citep{Barons2017} and nuclear emergency management \citep{Leonelli2013} applications this condition was broadly met. We suppose the centre consists of $m$ panels of experts denoted by $G_1,\dots,G_m$;
\item there must be a consensus about the policies $d$ that could be scrutinized and eventually implemented by the centre. In other words, all individuals in the centre must agree on a set $\mathbb{D}$ of decision rules whose efficacy might be examined by the IDSS. The choice of $\mathbb{D}$ is usually resolved using decision conferencing \citep{French2009} across panel representatives, users and stakeholders. We refer to this condition as \textit{policy consensus};
\item there must also be a consensus about the appropriate utility structure underlying a set of agreed attributes against which the efficacy of any policy is evaluated. So all individuals in the centre need to agree on the class $\mathbb{U}$ of utility functions supported by the IDSS. For instance, this consensus might be that the centre's utility function has utility independent attributes \citep{Keeney1976}. Again the choice of $\mathbb{U}$ is often resolved through decision conferencing. We refer to this condition as \textit{utility consensus};
\item consensus also needs to be found about an overarching description of the dynamics driving the process. We assume that  all panellists make their inferences in a parametric setting where a random vector $\bm{Y}$ is parametrised by a vector $\bm{\theta}$, where $\bm{Y}$ defines the variables of the process whilst $\bm{\theta}$ is the parameter vector on which inference is made. Then such a consensus consists of an agreement of all involved on the variables $\bm{Y}$, where, for each policy $d\in\mathbb{D}$, each utility function $u\in\mathbb{U}$ is a function of $\bm{Y}$ together with a set of qualitative statements about the dependence between various functions of $\bm{Y}$ and $\bm{\theta}$. This can take a variety of forms depending on the domain of application. In this paper we mainly focus on dependence structures represented by BNs, although  our methods apply equally well to other frameworks \citep{Smith2015}. We refer to this condition as \textit{structural consensus}.
\end{itemize}

The union of the policy, utility and structural consensus is called the \textit{common-knowledge class} (CK-class) and describes the agreement of all individuals on the components of the system and their relationships with each other. The CK-class defines the \textit{qualitative} structure of the domain investigated and therefore more easily provides a framework for the group's agreement \citep{Smith1996}. Protocols to guide the construction of the sets $\mathbb{D}$ and $\mathbb{U}$, and the identification of an overarching probabilistic model for the structural consensus have been recently defined \citep{Barons2017}.

Given this overarching qualitative structure has been agreed by the centre and represented by the CK-class, then an agreement on how to populate this class with \textit{quantitative} statements must be found. To this end, we assume the following condition holds:
\begin{itemize}
\item the centre must find a consensus about who is expert about what. In a formal sense, this implies that all panellists are prepared to adopt the beliefs of the designated expert panel in a specific sub-domain of the process as their own.
\end{itemize}
Thus in an IDSS beliefs' specifications are delegated to the most informed panel. Each  panel then, given a CK-class, individually delivers the necessary quantities for the computation of expected utilities concerning the variables under their jurisdiction.  However, as illustrated by influence of cost of oil and weather on food production and accessability to food, there is in general no guarantee that the individual beliefs of the panels can be combined to give a probabilistic coherent overall picture of the process. For the purposes of a formal Bayesian decision analysis an IDSS  needs to entertain the following property.

\begin{definition}
An IDSS is said to be \emph{adequate} for a CK-class if it can unambigously calculate the expected utility score of any decision $d\in\mathbb{D}$ and any utility function $u\in\mathbb{U}$ from the beliefs of the panels $G_1,\dots,G_m$.
\end{definition}

It is vital for an IDSS to be adequate since otherwise it could not produce a ranking of the available polices \citep{Keeney1976} and would therefore not be of any help to the decision centre for implementing and justifying any policy choice.  

\subsection{A real-world example}
\label{sec:example}
After a series of decision conferences with local authorities from Warwickshire County Council, stakeholders and potential decision makers, \citet{Barons2017} identified three areas that are impacted by increasing household food insecurity: health ($Y_1$), educational attainment ($Y_2$) and social cohesion ($Y_3$). Of course the cost ($Y_4$) associated to the enactment of any policy is deemed relevant in this domain. Measurable indices were developed for each of these areas - for instance, educational attainment is assessed by the percentage of pupils not failing a combination of UK school examinations. Further details about the form of the attributes are beyond the scope of this paper and we refer to \citet{Barons2017} for a discussion of these \citep[see also][]{Leonelli2017}.

\begin{figure}
\entrymodifiers={++[o][F-]}
\centerline{
\xymatrix{
\mbox{\large{$4$}}&\mbox{\large{$2$}}\ar[d]\\
\mbox{\large{$1$}}\ar[u]\ar[ur]\ar[r]&\mbox{\large{$3$}}
}
}
\caption{
BN representing the relationship between the four attributes in the food insecurity example, where the vertex $i$ is associated to the random variable $Y_i$, $i=1,\dots,4$. \label{fig:BN}}
\end{figure}
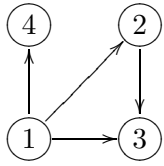

Notice that of course  in a reliable description of the food system any decision support system needs to account for the probabilistic dependence over a much larger vector of variables. But for the illustrative purposes of this example, we assume the decision centre agrees that these four random variables provide an overall, sufficient description of the household food system. The structural consensus of an IDSS with these four random variables then consists of a conditional independence structure that we suppose here to be depicted by the BN in Fig. \ref{fig:BN}. This states that given different levels of the health attribute, the associated costs are independent of both educational attainment and social cohesion.

Even in such a simple example we can highlight the heterogeneity of the food system and the consequent need of an IDSS. So for instance beliefs about the health attribute are delivered by doctors and public health experts; educational attainment is under the jurisdiction of school representatives and teachers; social unrest is overseen by sociologists, whilst judgements about costs are given by politicians and policymakers.

We consider a decision space $\mathbb{D}$ comprising of three possible policies: either an increase ($d_0$), a decrease ($d_1$) or not a change ($d_2$) of the number of pupils eligible for free school meals in Warwickshire. The UK government has already implemented this type of policy to give pupils a healthy start in life, since evidence seems to point towards an improvement of development and social skills of young children that eat a healthy meal together at lunchtime \citep{Kitchen2013}. We suppose henceforth that the decision centre, consisting of local authorities and stakeholders, agrees to consider only these three  policies. 

Lastly, the utility consensus might correspond to an agreement of a specific utility factorization over these four attributes. For instance, letting $\bm{y}=(y_1,y_2,y_3,y_4)$, where $y_i$ is an instantiation of $Y_i$, the centre might find an agreement that the utility function factorizes additively. Specifically,
\begin{equation}
u(d,\bm{y})=k_1(d)u_1(y_1,d)+k_2(d)u_2(y_2,d)+k_3(d)u_3(y_3,d)+k_4(d)u_4(y_4,d),
\label{eq:last}
\end{equation}
where $k_i(d)\in(0,1)$ and $d\in\mathbb{D}$. Given this agreed factorization, the specific form of the functions $u_i(y_i)$ is then elicited by the appropriate expert panel. 

In Section \ref{sec6} below we illustrate the methodology we introduce in this paper using this simple IDSS for household food security.

\subsection{Contributions}
\label{sec:contributions}
Recent advances in the integration of distributed expert judgements in complex systems have been reported in \citet{Leonelli2015} and \citet{Smith2015}. It has been demonstrated there that, perhaps surprisingly, it is common to be able to define a coherent system  by only specifying qualitative relationships between its random variables and quantifying a few of their associated summaries. Although some work has addressed the difficulties associated to the combination of expert judgments in multivariate systems \citep[e.g.][]{Faria1997,Farr2014}, none of these formally took into account the heterogeneity of the domain to be modelled.

In \citet{Smith2015} we  focused on the inferential full-distributional difficulties associated to this integration. However, a formal Bayesian decision analysis is  based on the maximization of an expected utility (EU) function and this often only depends on some simple summaries of key output variables, for example a few low order moments. By requesting from the relevant panels only the value of these expectations, the implementation of an IDSS can become orders of magnitude more manageable. Panels then just need to communicate a few summaries of their analysis: a trivial and fast task to perform within most inferential systems. In these cases real time decision support is thus feasible even when the system is huge.

 We demonstrate below that the EUs of such an IDSS are usually polynomials whose indeterminates are functions of the panels' delivered summaries. This polynomial structure enables us to identify new separation conditions, often implicit in standard conditional independence over the parameters of certain graphical models \citep{Freeman2011,Spiegelhalter1990} and milder than those of \citet{Smith2015},  sufficient to guarantee that an IDSS is adequate. An adequate IDSS is then capable of  supporting decision centres by providing a sound and coherent ranking of the available policies together with their associated EU scores.
 
Under the conditions derived above, we develop new propagation algorithms for BNs, here called \textit{algebraic substitutions}, for the distributed computations of an IDSS EU scores. These generalize the theory of the computation of moments of decomposable functions  \citep{Cowell1999,Nilsson2001} to multilinear ones.  Algebraic substitutions mirror the recursions of \citet{Lauritzen1992} for the computation of the first two moments of chain graph models. Here, focusing only on specific BN models, we are able to explicitly compute any joint moment and provide an intuitive graphical interpretation of the associated propagation rules.

Importantly, the recognition of the polynomial nature of EUs also enables us to analyze efficiently as well as exactly even large problems using software for symbolic manipulations (or \textit{computer algebra} software), e.g. Mathematica \citep{Mathematica}. Assuming the panels are able to deliver a vector of required summaries from the complex probabilistic model they plan to use, the software is then capable of combining them using algebraic substitutions to compute the associated EU scores almost instantaneously and in real-time to evaluate the candidate policies available to a decision centre. This is a critical property of any decision support system and in Section \ref{sec6} we give an illustration of how this can be achieved with computer algebra software.

\section{An algebraic description of IDSSs}
\label{section2}
We start by giving a polynomial description of the EUs of an IDSS. Consider a random vector $\bm{Y}=(\bm{Y}_i)_{i\in[m]}$, $[m]=\{1,\dots,m\}$,  where a subvector $\bm{Y}_i$ of $\bm{Y}$ is under the jurisdiction of a panel of experts $G_i$, $i\in[m]$. Let $\bm{y}\in\bm{\mathbb{Y}}$ and $\bm{y}_i\in\bm{\mathbb{Y}}_i$ be instantiations of $\bm{Y}$ and $\bm{Y}_i$, respectively. Assume each panel of experts delivers beliefs about $\bm{\theta}_i$, the parameter of the density $f_i$ over $\bm{Y}_i\;|\;(\bm{\theta}_i,d)$, where $d\in\mathbb{D}$ is one of the available policies in the decision space $\mathbb{D}$.  Suppose $\bm{\theta}_i$ takes values in $\bm{\Theta}_i$ and let $\bm{\theta}=(\bm{\theta}_i)_{i\in[m]}$ take values in $\bm{\Theta}$. Let $f$, $\pi_i$ and $\pi$ denote densities over $\bm{Y}\;|\;(\bm{\theta},d)$, $\bm{\theta}_i\;|\; d$ and $\bm{\theta}\;|\;d$, respectively. 

The IDSS processes the panels' judgements in order to calculate various statistics of an \textit{attribute vector}, usually some function of $\bm{Y}$. For simplicity and with no loss of generality we assume in this paper that attributes coincide with $\bm{Y}$.  For the purpose of a formal Bayesian analysis the IDSS computes the set of \textit{EU scores} $\{\bar{u}(d): d\in\mathcal{D}\}$ as a function of both the utility function $u(\bm{y},d)$ and the probability statements of the individual panels. The IDSS would then recommend to follow the policy $d^*$ with the highest EU score, $\bar{u}(d^*)$, where the EU is computed as
\begin{equation*}
\bar{u}(d)=\int_{\bm{\Theta}}\bar{u}(d\;|\; \bm{\theta})\pi(\bm{\theta}\;|\;d)\dr\bm{\theta},\\
\end{equation*}
and
\begin{equation*}
\bar{u}(d\;|\;\bm{\theta})=\int_{\bm{\mathbb{Y}}}u(\bm{y},d)f(\bm{y}\;|\; \bm{\theta},d)\dr \bm{y},
\end{equation*}
is the \textit{conditional expected utility (CEU)}.

By approaching the theory of IDSSs from an algebraic viewpoint, we are able to identify the necessary panels' summaries and the required assumptions for adequacy. In order to do this we first need to define the EU polynomials.

\begin{definition}
The CEU $\bar{u}(d\;|\;\bm{\theta})$ of an IDSS is called \emph{algebraic in the panels} if,  for each $d\in \mathcal D$ and for each panel $G_i$ in charge of $\bm{Y}_i$ with parameter $ \bm{\theta}_i$, $i\in[m]$, there exist functions $\bm{\lambda}_i(\bm{\theta}_i,d)$  of  $\bm{\theta}_i $ and $d$  such that $\bar u(d\;|\;\bm{\theta})$ is a square-free polynomial $q_d$ of the $\bm{\lambda}_i$
\[
\bar u(d\; |\; \bm{\theta}) = q_d\left( \bm{\lambda}_{1}(\bm{\theta}_{1},d),\cdots,\bm{\lambda}_m(\bm{\theta }_{m},d) \right).
\]
\end{definition}

Each $\bm{\lambda}_i$ is a vector of length $s_i$, where $s_i$ is the number of summaries each panel is required to deliver. Let $\bm{\lambda}_i(\bm{\theta}_i,d)=(\lambda _{ji}(\bm{\theta }_{i},d))_{j\in[s_i]}$, $[s_i]^0=[s_i]\cup\{0\}$  and $\bm{b}\in B=\times_{i\in[m]}[s_i]^0$ . For a given $\bm{b}=(b_i)_{i\in[m]}$  and $j\in[s_i]$, define  $b_{j,i}=0$ if $j\neq b_i$, $b_{j,i}=1$ if $j=b_i$ and $b_{0,i}=1$, for $i\in[m]$. It follows that  $b_{j,i}$ is not zero if and only if either $j=0$ or $j$ equals the $i$-th entry of $\bm{b}$. Let $\lambda_{0i}(\bm{\theta}_i,d)=1$, for every $\bm{\theta}_i\in \bm{\Theta}_i$, $d\in\mathcal{D}$ and $i\in[m]$.
 
\begin{example}
\label{ex:1}
Let $m=2$, $s_1=s_2=1$, i.e. there are two panels each delivering one summary only. Then $B=\{(0,0), (0,1), (1,0), (1,1)\}$. For  $\bm{b}=(0,1)\in B$ we have that $b_{0,1}=1$, $b_{1,1}=0$, $b_{0,2}=1$ and $b_{1,2}=1$.
\end{example}
\begin{definition}
The CEU  $\bar{u}(d\;|\;\bm{\theta})$ of an IDSS is called \emph{algebraic} if, for each $d\in \mathcal D$, $q_d$ is a square-free polynomial of the $\lambda_{ji}$, $i\in[m]$, $j\in[s_i]^0$, such that
\begin{equation}
q_{d}\left( \bm{\lambda }_{1}(\bm{\theta }_{1},d),\dots,\bm{\lambda }_{m}(\bm{\theta }_{m},d)\right)  = \sum_{\bm{b}\in B}k_{\bm{b}}(d)\lambda _{\bm{b}}(\bm{ \theta },d),  \label{algebraic utility} 
\end{equation}
with $k_{\bm{b}}(d)\in\mathbb{R}$ and
\begin{equation*}
 \lambda _{\bm{b}}(\bm{\theta },d) = \prod_{i\in[m]}\prod_{j\in[s_i]^0}\lambda_{ji}(\bm{\theta }_{i},d)^{b_{j,i}}. \label{eq:aceu2}
\end{equation*}
\end{definition}
Thus, $\lambda_{\bm{b}}$ is a monomial having at most one term not unity delivered by each panel and $k_{\bm{b}}(d)$ is a weight. 
\begin{example}
\label{ex:2}
Let the CK-class specify that $\bm{Y}=(Y_i)_{i\in[m]}$, where each variable $Y_i$ is binary and overseen by panel $G_i$. Assume that for all decisions $d\in \mathcal{D}$, $\theta_i=\mathbb{P}(Y_i|\theta_i,d)$, $\bm{\theta}=(\theta_i)_{i\in[m]}$, and that the CK-class includes the belief that $\independent_{i\in [m]} Y_i|\bm{\theta},d$, where $\independent$ denotes conditional independence \citep{Dawid1979}. Suppose the utility consensus consists of a utility factorization of the form 
\[
u(\bm{y})=u(y_1,\dots,y_m)=\sum_{i\in [m]}k_i(d)y_i+\sum_{i\in[m]}\sum_{i<j\leq m}k_{ij}(d)y_iy_j,
\]
where $k_i(d)$ and $k_{ij}(d)$ are jointly agreed criterion weights \citep{French2009}. With no further assumption the CEU can be written as
\begin{equation}
\bar{u}(d|\bm{\theta})=\sum_{i\in [m]}k_i(d)\lambda_{1i}(\theta_i,d)+\sum_{i\in[m]}\sum_{i<j\leq m}k_{ij}(d)\lambda_{1i}(\theta_i,d)\lambda_{1j}(\theta_i,d),
\label{eq:ex1}
\end{equation}
where $\lambda_{1i}(\theta_i,d)=\theta_i$. Thus equation (\ref{eq:ex1}) is an algebraic CEU and $\bm{\lambda}_i(\bm{\theta}_i,d)=(1,\theta_i)$.
\end{example}

\begin{example}
\label{ex:3}
A more complex example is given by two dependent continuous random variables $Y_1$ and $Y_2$ such that $\E(Y_1)=\theta_{01}$, $\V(Y_1)=\psi_1$, $\E(Y_2|Y_1=y_1)=\theta_{02}+\theta_{12}y_1$ and $\V(Y_2|Y_1=y_1)=\V(Y_2)=\psi_2$. Here $\mathbb{E}$ stands for expectation and $\mathbb{V}$ for variance. Assume the utility consensus includes an additive utility factorization $u(y_1,y_2,d)=k_1(d)u(y_1,d)+k_2(d)u(y_2,d)$, where each marginal utility function is in the family of quadratic utility functions \citep{Wakker2008}, i.e. $u(y_i,d)=a_i(d)y_i-b_i(d)y_i^2$, where $a_i(d)\in\mathbb{R}$ and $b_i(d)\in\mathbb{R}_{>0}$, $d\in\mathcal{D}$ and $i\in[2]$. Using standard properties of conditional moments \citep[see e.g.][]{Brillinger1969}, by leaving implicit the dependence on $d$ the CEU can be written as 
\begin{multline*}
\bar{u}(d|\bm{\theta})=k_1(a_1\theta_{01}-b_1\theta_{01}^2-b_1\psi_1)+\\
k_2(a_2\theta_{02}+a_{2}\theta_{12}\theta_{01}-b_2\theta_{02}^2-b_2\theta_{12}^2\theta_{01}^2-b_2\theta_{12}^2\psi_1-b_2\psi_2-2b_2\theta_{02}\theta_{12}\theta_{01}).
\end{multline*}
In this second example the CEU is again algebraic and 
\begin{align*}
\bm{\lambda}_1(\bm{\theta}_1,d)&=(1,a_1\theta_{01},b_1\theta_{01}^2,b_1\psi_1,\theta_{01},\theta_{01}^2,\psi_1),\\
\bm{\lambda}_2(\bm{\theta}_2,d)&=(1,a_2\theta_{02},a_2\theta_{12},b_2\theta_{02}^2,b_2\theta_{12}^2,b_2\psi_2,2b_2\theta_{02}\theta_{12}).
\end{align*}
\end{example}

To achieve adequacy we need the following property.
\begin{definition}
Let $
\mu _{ji}(d)= \mathbb{E}\left( \lambda_{ji}(\bm{\theta }_{i},d)^{b_{j,i}}\right)$, 
for a given $\bm{b}\in B$. We call an IDSS \emph{score separable} if, in the notation above, all panellists agree that, for all decisions $d\in \mathcal{D}$ and all indices $\bm{b}\in B$ such that $k_{\bm{b}}(d)\neq 0$,
\begin{equation*}
\mathbb{E}\left( \lambda _{\bm{b}}(\bm{\theta },d)\right)=\prod_{i\in[m]}\prod_{j\in[s_i]^0}\mu_{ji}(d). \label{uncor}
\end{equation*}
\label{def:scoresep}
\end{definition}

A score separable IDSS can then determine the EU score of any policy $d\in\mathbb{D}$ from the summaries $\mu_{ij}(d)$ individually delivered by the panels, $i\in[m]$, $j\in[s_i]$. This implies adequacy as formalized in Lemma \ref{lemma:ciao} below. For every $d\in\mathcal{D}$, let $\bm{\mu }_{i}(d)=\left( \mu _{ji}(d)\right)_{j\in[s_i]}$. 

\begin{lemma}
\label{lemma:ciao}
Suppose $G_{i}$ delivers its vectors of expectations $\bm{\mu }_{i}(d)$, $i\in[m]$, $d\in\mathcal{D}$. For an algebraic CEU, if the IDSS is score separable then it is adequate.
\end{lemma}

The proof of this result  follows from the definition of algebraic CEU in equation (\ref{algebraic utility}) and the definition of score separability.

\begin{example}[Example \ref{ex:2} continued]
From equation (\ref{eq:ex1}) we can deduce that the score separability condition corresponds to the factorization of the expectations $\E(\theta_{i}\theta_{j})$, $i,j\in[m]$, $i\neq j$, into $\E(\theta_i)\E(\theta_j)$.
\end{example}

\begin{example}[Example \ref{ex:3} continued]
\label{ex:5}
Score separability corresponds to the conditions
\[
\begin{array}{lcccr}
\E(\theta_{01}\theta_{12})=\E(\theta_{01})\E(\theta_{12})&&&&\E(\theta_{12}^2\theta_{01}^2)=\E(\theta_{12}^2)\E(\theta_{01}^2)\\
 \E(\theta_{12}^2\psi_1)=\E(\theta_{12}^2)\E(\psi_1)&&&&\E(\theta_{02}\theta_{12}\theta_{01})=\E(\theta_{02}\theta_{12})\E(\theta_{01}).
\end{array}
\]
\end{example}

\section{Moment and quasi independence}
\label{section3}

 Lemma \ref{lemma:ciao} shows that adequacy is guaranteed whenever score separability holds for algebraic CEUs. This implies that the expectation of certain functions of the panels' parameters separate appropriately. We first introduce conditions that ensure this type of separability and then in Section \ref{section4} identify classes of models that give rise to algebraic CEUs.

\begin{definition}
\label{def:partial}
Let $q_{d}(\bm{\lambda }_{_{1}}(\bm{\theta }_{1},d),\dots,\bm{\lambda }_{m}(\bm{\theta }_{m},d))$ be the algebraic CEU of an IDSS. An IDSS is called \emph{quasi independent} if 
\begin{equation*}
\E(q_{d}(\bm{\lambda }_{_{1}}(\bm{\theta }_{1},d),\dots,\bm{\lambda }_{m}(\bm{\theta }_{m},d)))=q_{d}(\E(\bm{\lambda }_{_{1}}(\bm{\theta }_{1},d)),\dots,\E(\bm{\lambda }_{m}(\bm{\theta }_{m},d))).
\end{equation*}
\end{definition}
This condition requires the expectation of the product of certain functions of the parameters overseen by different panels to be equal to the product of the individual expectations.

Often the $\lambda_{ji}$, $i\in[m]$, $j\in[s_i]$, are monomial functions of the panels' parameters. This was the case in Examples \ref{ex:2} and \ref{ex:3} above. It is therefore helpful to introduce the following independence condition specific for monomial functions. Let $<_{lex}$ denote a lexicographic order \citep{Cox2007}.

\begin{definition}
\label{def:moment}
Let $\bm{\theta}=(\theta_i)_{i\in[m]}\in\mathbb{R}^{m}$ be a parameter vector and $\bm{c}=(c_i)_{i\in[n]}\in\mathbb{Z}^n_{\geq 0}$. We say that $\bm{\theta}$ entertains \emph{moment independence} of order $\bm{c}$ if for any $\bm{a}=\left(a_i\right)_{i\in[n]}<_{lex} \bm{c}$, $\bm{a}\in\mathbb{Z}_{\geq 0}^n$, and letting $\bm{\theta}^{\bm{a}}=\theta_1^{a_1}\cdots\theta_n^{a_n}$, it holds
\begin{equation*}
\E\left(\bm{\theta}^{\bm{a}}\right)=\prod_{i\in[n]}\E\left(\theta_i^{a_i}\right).
\end{equation*}
\end{definition}

\begin{example}
In Example \ref{ex:5} we say that score separability holds for the EU of Example \ref{ex:3} if e.g. $\E(\theta_{01}\theta_{12})=\E(\theta_{01})\E(\theta_{12})$ and $\E(\theta_{01}^2\theta_{12}^2)=\E(\theta_{01}^2)\E(\theta_{12}^2)$. The first requirement correspond to a moment independence of degree $(1,1)$, whilst the second is a moment independence of degree $(2,2)$.
\end{example}

It is generally well known that standard probabilistic independence only guarantees that the first moment of a product can be written as the product of the moments. Separations for higher orders are implied by standard independence only through a cumulant parametrization, where the cumulant generating function for a product of independent random variables (defined as a random sum of independent realizations) is the composition of the respective cumulant generating functions \citep{Brillinger1969}.

For the purpose of decision support it is helpful to study moments, since expected utilities often formally depend on these. Consider for instance two parameters $\theta_1$ and $\theta_2$. Assume a CEU is equal to $\theta_1^2\theta_2^2$ and that a moment independence of order $(2,2)$ holds. Then
\begin{eqnarray}
\mathbb{E}\left(\theta_1^2\theta_2^2\right)&=&\mathbb{E}\left(\theta_1^2\right)\mathbb{E}\left(\theta_2^2\right)\nonumber\\
&=&\mathbb{E}(\theta_1)^2\mathbb{E}(\theta_2)^2+\mathbb{E}(\theta_1)^2\mathbb{V}(\theta_2)+\mathbb{E}(\theta_2)^2\mathbb{V}(\theta_1)+\mathbb{V}(\theta_1)\mathbb{V}(\theta_2).
\label{eq:towerex}
\end{eqnarray}
The same expression is obtained when using sequentially the tower rule of expectations and the law of total variance under the assumption of independence of the two parameters above \citep{Brillinger1969}. Therefore, the expression obtained under moment independence is reasonable and coincides with the one implied by the independence of $\theta_1$ and $\theta_2$. However the condition we need for equation (\ref{eq:towerex}) to hold \textit{does not require} $\theta_1$ and $\theta_2$ to be independent.

\section{Adequate combinations of probabilistic outputs}
\label{section4}

 Given the above definitions  of new independence concepts tailored to IDSSs, we can now study situations where these can be shown to be adequate and therefore provide a coherent, operational support tool to decision centres. 
 
\begin{proposition}
\label{lemma:ciaociao}
Let $q_{d}(\bm{\lambda }_{_{1}}(\bm{\theta }_{1},d),\dots,\bm{\lambda }_{m}(\bm{\theta }_{m},d))$ be an algebraic CEU of a quasi independent IDSS. The IDSS is adequate if panel $G_i$ delivers the vectors of expectations $\bm{\mu}_i(d)$, for all $i\in[m]$ and $d\in\mathcal{D}$.
\end{proposition}

This result follows by noting that quasi independence implies score separability since
\[
\bar{u}(d)=q_d(\E(\bm{\lambda}_1(\bm{\theta}_1,d)),\dots,\E(\bm{\lambda}_m(\bm{\theta}_m,d)))=\sum_{\bm{b}\in B}k_{\bm{b}}(d)\prod_{i\in[m]}\prod_{j\in[s_i]^0}\mu_{ji}(d).
\]

Assuming the CEU is a polynomial in the panels' parameters, under a specific moment independence assumption we have a more operational result. 

\begin{corollary}
\label{lemma:ciao3}
Let $q_{d}(\bm{\lambda }_{_{1}}(\bm{\theta }_{1},d),\dots,\bm{\lambda }_{m}(\bm{\theta }_{m},d))$ be an algebraic CEU of an IDSS, $\bm{\theta}_i=(\theta_{ji})_{j\in[s_i]}$ and $\lambda_{ji}(\bm{\theta}_i,d)=\bm{\theta}_{i}^{\bm{a}_{ji}}$, with $\bm{a}_{ji}\in\mathbb{Z}_{\geq 0}^{s_i}$, $i\in[m]$, $j\in[s_i]$. Let $\bm{a}^*_i=({a_{ji}^*})_{j\in[s_i]}$, where $a_{ji}^*$ is the greatest element in $\{a_{ji}:j\in[s_i]\}$, $i\in[m]$, and let ${\bm{a}^*}=({\bm{a}_i^*})_{i\in[m]}$. Let $\bm{\theta}=(\bm{\theta}_i)_{i\in[m]}$ and assume the CK-class includes a moment independence assumption of order $\bm{a}^*$. The IDSS is then adequate if panel $G_i$ delivers  the vectors of expectations $\bm{\mu}_i(d)$, for all  $i\in[m]$ and  $d\in\mathcal{D}$.
\end{corollary} 

The proof of this result is given in Appendix \ref{app:A}. Proposition \ref{lemma:ciaociao} and Corollary \ref{lemma:ciao3} formalize the independence conditions required for an IDSS to be adequate, under the assumption of an algebraic CEU.  In practice it is often the case that a CEU is algebraic  \citep[e.g.][and in Examples \ref{ex:2} and \ref{ex:3}]{Madsen2005}. However, there are particular families of utility factorizations and statistical models that \emph{ensure} the associated CEU is algebraic. We define these classes below and prove that their associated CEU is algebraic.

\begin{definition}
\label{def:panelsep}
Let $\bm{Y}_i$ be the vector overseen by panel $G_i$, $i\in[m]$.  A  utility function over $\bm{Y}_1,\dots,\bm{Y}_m$ is called \emph{panel separable} if it factorizes as
\[
u(\bm{y}_1,\dots,\bm{y}_m,d)=\sum_{I\in{\mathcal{P}_0([m])}}k_{I}(d)\prod_{i\in I}u_i(\bm{y}_i,d),
\]   
where $\mathcal{P}_0$ is the power set without the empty set and $k_I(d)$ is a criterion weight. 
\end{definition}

\begin{definition}
Under the conditions of Definition \ref{def:panelsep}, a utility function over $\bm{Y}_1,\dots, \bm{Y}_m$ is called \emph{additive panel separable} if it factorizes as
\[
u(\bm{y}_1,\dots,\bm{y}_m,d)=\sum_{i\in[m]}k_i(d)u_i(\bm{y}_i,d).
\]
\end{definition}

Under the assumption of an (additive) panel separable utility, each panel can model its preferences over the variables under its jurisdiction using a marginal utility function of its choice. A large class of utilities, often used in practice,  are \textit{polynomial} \citep{Muller1987}. For simplicity, we here consider only the case when marginal utility functions have univariate arguments. 

\begin{definition}
A \emph{polynomial} utility function over $y_i$ of degree $n_i$ is defined as
\[
u(y_i,d)=\sum_{j\in[n_i]}\rho_{ij}(d)y_i^{j},
\]
where the coefficients $\rho_{ij}(d)\in\mathbb{R}$ and the domain of $y_i$ need to entertain some constraints.\footnote{For simplicity, we assume the intercept to be equal to zero since utilities are unique up to positive affine transformations.} 
\end{definition} 
An explicit derivation of the required constraints can be found in \citet{Keeney1976} and \citet{Muller1987}.

The probabilistic model class we consider here is a specific structural equation model (SEM) \citep{Bowen2011,Wall2000}, where each variable is defined through a polynomial function. Henceforth we call these a \textit{polynomial SEM}. SEMs were first introduced as a modelling approach in  the social sciences \citep{Westland2015} and are nowadays widely used especially in the causal literature \citep{Pearl2000}. 

\begin{definition}
\label{def:polystrut}
Let $\bm{Y}=(Y_i)_{i\in[m]}$ be a random vector. A polynomial SEM is defined by
\begin{equation*}
Y_i=\sum_{\bm{a}_i\in A_i}\theta_{i\bm{a}_i}\bm{Y}_{[i-1]}^{\bm{a}_i}+\varepsilon_i,\hspace{1.5cm} i\in[m],
\end{equation*}
where $A_i\subset \mathbb{Z}^{i-1}_{\geq 0}$, $\varepsilon_i$ is a random error with mean zero and variance $\psi_i$, $\theta_{i\bm{a}_i}$ is a parameter, $i\in[m]$, $\bm{a}_i\in A_i$, and $\bm{Y}_{[i{-1}]}=(Y_j)_{j\in[i{-1}]}$, with $[0]=\emptyset$.
\end{definition}
An alternative formulation of a polynomial SEM in terms of distributions is
\begin{equation*}
Y_i\;|\;(\bm{\theta}_i,\bm{Y}_{[i-1]})\sim \left(\sum_{\bm{a}_i\in A_i}\theta_{i\bm{a}_i}Y_{[i-1]}^{\bm{a}_i},\psi_i\right),
\end{equation*}
where $\bm{\theta}_i=(\theta_{i\bm{a}_i})_{\bm{a}_i\in A_i}$ and $i\in[m]$. These models are suitable candidates for a CK-class since their definition is qualitative in nature and requires only the specification of the relationships between the random variables together with a few selected moments.
 
For polynomial SEMs and panel separable utilities the following holds.

\begin{theorem}
\label{theo:ciao}
Assume panel $G_i$ is responsible for $Y_i$, $i\in[m]$ and that the CK-class of an IDSS includes a panel separable utility and a polynomial SEM. Assume also that each panel agreed to model its marginal utility with a polynomial utility function. Then, under quasi independence, the IDSS is score separable.
\end{theorem}

The proof of this theorem is given in Appendix \ref{app:B}. Theorem \ref{theo:ciao} together with Lemma \ref{lemma:ciao} shows that IDSSs, whose CK-class includes polynomial SEMs and panel separable utilities, can uniquely compute EU scores  from the individual judgements of the panels. By construction, the quasi independence condition of Theorem \ref{theo:ciao} actually corresponds  to a moment independence. The order of such independence depends on the polynomial form of both the SEM and the utility function. In Section \ref{section5} we  identify the order of the moment independence condition required for adequacy in a subclass of polynomial SEMs.

\section{Bayesian networks}
\label{section5}
The subclass of polynomial SEMs we study next consists of BN models where each vertex is defined by a linear regression over its parents. For this model class we are able to deduce the exact moment independence required for adequacy. 

\begin{definition}
\label{def:lin}
A BN over a directed acyclic graph (DAG) $\Gr$ with vertex set $V(\Gr)=\{i:i\in[m]\}$ and edge set $E(\Gr)$ is  a \emph{linear SEM} if each variable $Y_i$ is defined as
\begin{equation}
\label{eq:def}
Y_i=\theta_{0i}+\sum_{j\in \Pi_i} \theta_{ji}Y_j+\varepsilon_i,
\end{equation}
where $\Pi_i$ is the parent set of $i$ in $\Gr$, $\varepsilon_i$ is a random error with mean zero and variance $\psi_i$ and $\theta_{0i},\theta_{ji}\in\mathbb{R}$.
\end{definition} 
 Although such models are often multivariate Gaussian, in general this does not need to be the case.

As \citet{Sullivant2010}, we consider regression parameters as indeterminates in a polynomial function. We associate these to edges and vertices of the underlying DAG. For $i\in [m]$, let $\theta_{0i}'=\theta_{0i}+\varepsilon_{i}$ be the indeterminate associated to the vertex $i$, whilst $\theta_{ij}$ is associated to the edge $(i,j)\in E(\Gr)$.\footnote{We think of $\theta_{0i}'$ as a parameter although this consists of the sum of a parameter $\theta_{01}$ and an error $\varepsilon_i$. Note however that from a Bayesian viewpoint these are  both random variables.} Define $\mathbb{P}_i$ to be the set of rooted paths in $\Gr$ ending in $Y_i$. A \textit{rooted path} of length $n+1$ from $i_1$ to $j_n$ is a sequence comprising of a vertex in $V(\Gr)$ and $n$ distinct edges in $E(\Gr)$ is such that
$
(i_1,(i_1,j_1),\dots, (i_k,j_k),(i_{k+1},j_{k+1}),\dots,(i_n,j_n)),
$
 where $j_k=i_{k+1}$, $k\in[n{-1}]$, $i_k,j_k\in[m]$. For every element $P\in\mathbb{P}_i$ we define $\bm{\theta}_P$ as
 \begin{equation*}
 \bm{\theta}_P=\prod_{i\in P}\theta_{0i}'\prod_{(i,j)\in P}\theta_{ij},
 \end{equation*}
and, as \citet{Sullivant2010}, we call $\bm{\theta}_{P}$ the \textit{path monomial}.

\begin{example}
Consider the DAG in Figure \ref{fig:BN} associated to the food security example. For instance, the set $\mathbb{P}_3$ is equal to
\begin{equation*}
\label{eq:paths}
\{(3),(2,(2,3)), (1,(1,3)), (1,(1,2),(2,3))\},
\end{equation*}
and $\theta_{03}'$, $\theta_{02}'\theta_{23}$, $\theta_{01}'\theta_{13}$ and $\theta_{01}'\theta_{12}\theta_{23}$ are the corresponding path monomials.
\end{example}

We call \textit{algebraic substitution} the process of substituting the linear regression expression  of a random variable of the DAG, as in equation (\ref{eq:def}), into the one of the child variable. An example illustrates this process.

\begin{example}
For the DAG in Figure \ref{fig:BN}, a linear SEM is defined as
\begin{equation}
\label{eq:sem}
\begin{array}{lcccccccl}
Y_4=\theta_{04}+\theta_{14}Y_1+\varepsilon_1, &&&&&&&& Y_3=\theta_{03}+\theta_{13}Y_1+\theta_{23}Y_2+\varepsilon_3,\\
Y_2=\theta_{02}+\theta_{12}Y_1+\varepsilon_2,&&&&&&&& Y_1=\theta_{01}+\varepsilon_1.
\end{array}
\end{equation}
An algebraic substitution of the variables in the definition of $Y_3$ entails
\begin{align*}
Y_3&=\theta_{03}+\theta_{13}(\theta_{01}+\varepsilon_1)+\theta_{23}(\theta_{02}+\theta_{12}Y_1+\varepsilon_2)+\varepsilon_3\\
&=\theta_{03}'+\theta_{13}\theta_{01}'+\theta_{23}\theta_{02}'+\theta_{23}\theta_{12}Y_1.
\end{align*}
The additional algebraic substitution of $Y_1$ gives 
\begin{equation}
Y_3=\theta_{03}'+\theta_{13}\theta_{01}'+\theta_{23}\theta_{02}'+\theta_{23}\theta_{12}\theta_{01}'.\label{eq:y3'}
\end{equation}
\end{example}
It is of special interest that after this substitution $Y_3$ is now uniquely defined in equation (\ref{eq:y3'}) in terms of path monomials. Proposition \ref{prop:algsub} formalizes that this occurs for any variable of a DAG defined as a linear SEM and links algebraic substitutions to conditional expectation operators.

\begin{proposition}
\label{prop:algsub}
For a linear SEM over a DAG $\Gr$, through algebraic substitutions each variable $Y_i$, $i\in[m]$, can be written as
\begin{equation}
\label{eq:1}
Y_i=\sum_{P\in\mathbb{P}_i}\bm{\theta}_{P},
\end{equation}
and letting  $\bm{\theta}_{i}=(\theta_{0i}',\theta_{ji})_{j\in \Pi_i}$ and $\bm{\theta}=(\bm{\theta}_i)_{i\in[m]}$, $i\in[m]$, we then have that 
\begin{equation}
\label{eq:2}
\E(Y_i\;|\;\bm{\theta},d)=\sum_{P\in\mathbb{P}_i}\bm{\theta}_P.
\end{equation}
\end{proposition}
The proof of this result is given in Appendix \ref{app:C}.

\subsection{Additive factorizations.}
\label{sec:add}
By Proposition \ref{prop:algsub}, the CEU of polynomial additive panel separable utilities can be written as a polynomial function of a set of monomials readable into the structure of the DAG.

\begin{lemma}
\label{lemma:addeu}
Consider a linear SEM over a DAG $\Gr$. Assume that  $u(\bm{y})$ can be written as
\begin{equation*}
u_i(\bm{y})=\sum_{i\in[m]}k_i(d)u_i(y_i).
\end{equation*}
and that $u_i$ is a polynomial utility function of degree $n_i$. Then the CEU is algebraic and can be written as
\begin{equation}
\label{eq:addeu}
\bar{u}(d\;|\;\bm{\theta})=\sum_{i\in[m]}k_i(d)\sum_{j\in[n_i]}\rho_{ij}(d)\sum_{|\bm{a}_i|=j}\binom{j}{\bm{a}_i}\bm{\theta}_{\mathbb{P}_i}^{\bm{a}_i},
\end{equation}
where $\bm{a}_i=(a_{ij})_{j\in[\#\mathbb{P}_i]}\in\mathbb{Z}_{\geq 0}^{\#\mathbb{P}_i}$, $\bm{\theta}_{\mathbb{P}_i}=\prod_{P\in\mathbb{P}_i}\bm{\theta}_P$, $\binom{j}{\bm{a}_i}$ is a multinomial coefficient, $\#\mathbb{P}_i$ is the number of elements in $\mathbb{P}_i$ and $|\bm{a}_i|=\sum_{j\in\#\mathbb{P}_i}a_{ij}$.
\end{lemma}
This result follows by noting, from Equation (\ref{eq:2}),  that the CEU equals
\begin{equation*}
\label{eq:additiveEU}
\E(\bar{u}(d\;|\;\bm{\theta}))=\sum_{i\in[m]}k_i(d)\sum_{j\in[n_i]}\rho_{ij}(d)\Big(\sum_{P\in\mathbb{P}_i}\bm{\theta}_P\Big)^j,
\end{equation*}
and then applying the Multinomial Theorem \citep{Cox2007}.

Equation (\ref{eq:addeu}) is an instance of the computation of the moments of a decomposable function as studied in \citet{Cowell1999} and \citet{Nilsson2001}. In Lemma \ref{lemma:addeu} we explicitly deduce the required monomials and their degree and in Section \ref{sec:multi} we generalise these results  to multilinear functions. 

Lemma \ref{lemma:addeu} has an appealing intuitive graphical interpretation which is particularly useful for the computation of the EU's monomials. The $j$-th non central moment of any  $Y_i$ can be written as the sum of the monomials $\bm{\theta}_{\mathbb{P}_i}$ with degree $j$. By the properties of multinomial coefficients, this sum can be thought of as the sum over the set of unordered $j$-tuples of rooted paths ending in $Y_i$. Let $\mathbb{P}^j_i$ be the set of unordered $j$-tuples from $\mathbb{P}_i$. For a $P\in\mathbb{P}^j_i$, the multinomial coefficient in equation (\ref{eq:addeu}) counts the distinct permutations of the elements of $P$, denoted as $n_{P_i}$. We then have that, 
\begin{equation}
\label{eq:grapheu}
\sum_{|\bm{a}_i|=j}\binom{j}{\bm{a}_i}\bm{\theta}_{\mathbb{P}_i}^{\bm{a}_i}=\sum_{P\in\mathbb{P}^j_i}n_{P_i}\prod_{p\in P}\bm{\theta}_p.
\end{equation} 
 Equation (\ref{eq:grapheu}) shows an intuitive graphical interpretation of equation (\ref{eq:addeu}), as illustrated in the following example.

\begin{example}
For the vertex $4$ in the DAG of Figure \ref{fig:BN} the set $\mathbb{P}_4$ is equal to $\{(4), (1,(1,4))\}$. From the left hand side of equation (\ref{eq:grapheu}), $Y_4^2$ can be written as 
\begin{equation}
\label{eq:exbn1}
\theta'^2_{04}+\theta'^2_{01}\theta_{14}^2+2\theta_{01}'\theta_{14}\theta_{04}'.
\end{equation}
This polynomial can be also deduced by simply looking at the DAG. Note that 
\begin{equation*}
\mathbb{P}_4^2=\Big\{\big((4),(4)\big), \big((1,(1,4)),(1,(1,4))\big), \big((4),(1,(1,4))\big)\Big\}.
\end{equation*}
The first and second monomial in equation (\ref{eq:exbn1}) correspond to the first and second element of $\mathbb{P}_4^2$ respectively, whilst the last elements of this set, having two distinct permutations, is associated to the third monomial in equation (\ref{eq:exbn1}).
\end{example}

From Lemma \ref{lemma:addeu} we can deduce the independences needed for adequacy in IDSSs whose CK class includes a BN defined as a linear SEM. Note that $\bm{\theta}_{\mathbb{P}_i}$, defined as $\prod_{P\in\mathbb{P}_i}\bm{\theta}_P$, might include multiple times the same parameter, $\theta$ say, if $\theta$ is associated to a vertex/edge appearing in different paths ending in $Y_i$. We let $\bm{\theta}_{\Gr_i}$ be the simplified version of $\bm{\theta}_{\mathbb{P}_i}$ where each parameter appears only once and $\bm{\theta}_{\Gr_i}^{\bm{c}_i}$ is the simplified version of $\bm{\theta}_{\mathbb{P}_i}^{\bm{a}_i}$ where each element of $\bm{c}_i$ equals the sum of the $a_{ij}$ associated to the same parameter.  Let $r_i$ be the number of distinct parameters in $\bm{\theta}_{\mathbb{P}_i}$.

\begin{theorem}
\label{theo:pr}
Suppose  the CK-class of an IDSS includes a linear SEM over a DAG $\Gr$, where panel $G_i$ oversees  $Y_i$, $i\in[m]$, and an additive panel separable utility function. Suppose panel $G_i$ agreed to use a polynomial utility function of degree $n_i$, $i\in[m]$. If $\bm{\theta}_{\Gr_i}$ entertains moment independence of order $\bm{c}_i$ for every $\bm{c}_i\in\mathbb{Z}_{\geq 0}^{r_i}$ such that $|\bm{c}_i|=n_i$ and $i\in[m]$, then the IDSS is adequate.  
\end{theorem}

The proof of this result is given in Appendix \ref{app:E}.
Theorem \ref{theo:pr} gives the specific moment independences necessary for the IDSS's adequacy. By requesting the collective to agree on these independences, the IDSS can then quickly produce a unique EU score for each policy. Panels are informed on the summaries they need to deliver to the IDSS since these are the only quantities of which the EU is a function. An illustration of the usefulness of this result in practice is given in Section \ref{sec6}.

\subsection{Multilinear factorizations.}
\label{sec:multi}
By approaching the combination of outputs in BN models from an algebraic viewpoint, we are able to generalize in a straightforward manner the results in Section \ref{sec:add} about additive/decomposable factorizations so that they apply to multilinear functions. Let  $\#\mathbb{P}_{i}=m_i$, i.e. there are $m_i$ rooted paths ending in $Y_i$. Let $\bm{l}_i=(l_{ij})_{j\in[m_i]}\in\mathbb{Z}^{m_i}_{\geq 0}$ be the vector listing the lengths of such paths and  $\bm{l}=(\bm{l}_i)_{i\in[m]}$.  For a vector $\bm{a}=(a_i)_{i\in[m]}\in\mathbb{Z}^m$, we write $\bm{l}\simeq \bm{a}$ if both $|\bm{a}|=|\bm{l}|$ and, for all  $i\in[m]$, $|\bm{l}_i|=a_i$.

\begin{lemma}
\label{lemma:multieu}
For a linear SEM over a DAG $\Gr$, suppose  the utility function $u(\bm{y},d)$ can be written 
\begin{equation*}
\label{eq:multidag}
u(\bm{y},d)=\sum_{I\in\mathcal{P}_0([m])}k_I(d)\prod_{i\in I}u_i(y_i,d).
\end{equation*} 
Now suppose $u_i$ is a polynomial utility function of degree $n_i$, $\bm{n}=(n_i)_{i\in[m]}$,  $i\in[m]$ and $\bm{0}$ is a vector of dimension $m$ with only zero entries. The CEU is then algebraic and can be written as
\begin{equation}
\label{eq:multieu}
\bar{u}(d\;|\;\bm{\theta})=\sum_{\bm{0}<_{lex}\bm{a}\leq_{lex}\bm{n}}c_{\bm{a}}(d)\sum_{\bm{l}\simeq \bm{a}}\binom{|\bm{a}|}{\bm{l}}\bm{\theta}_{\mathbb{P}}^{\bm{l}},
\end{equation}
where $c_{\bm{a}}(d)=k_J(d)\prod_{j\in J}\rho_{ja_j}(d)$, $J=\{j\in[m]: a_j\neq 0\}$, and $\bm{\theta}_{\mathbb{P}}=\prod_{i\in[m]}\bm{\theta}_{\mathbb{P}_i}$.
\end{lemma}

The proof of Lemma \ref{lemma:multieu} is given in Appendix \ref{app:F} Lemma \ref{lemma:multieu} makes a significant generalization to the theory of the computation of moments in decomposable/additive functions of \citet{Cowell1999} and \citet{Nilsson2001} extending these well-known formulae so that they apply in the much wider context of multilinear functions of BNs defined as linear SEMs.  It is interesting to note that Lemma \ref{lemma:multieu} is connected to the propagation algorithms first developed in \citet{Lauritzen1992} to compute the first two moments of certain chain graphs. Here, focusing on a specific class of continuous DAG models, we are able to explicitly compute, through algebraic substitution, not only the first two moments, but also any other higher order moment of the distribution associated with the graph.  

Using again the properties of multinomial coefficients, we can relate  equation (\ref{eq:multieu}) to the topology of the graph and its rooted paths. For an $\bm{a}\in\mathbb{Z}^{m}_{\geq 0}$, let $\mathbb{P}_{\bm{a}}=\times_{a_i\neq 0} \mathbb{P}^{a_i}_i$, where $\times$ denotes the Cartesian product. This set consists of the unordered $|\bm{a}|$-tuples of rooted paths, where in each tuple there are $a_i$ paths ending at $Y_i$. For each element $P\in\mathbb{P}_{\bm{a}}$, let $n_P=\sum_{a_i\neq 0}n_{P_i}$. Then we have that, following the same reasoning outlined for additive factorizations,
 \begin{equation*}
 \sum_{\bm{l}\simeq \bm{a}}\binom{|\bm{a}|}{\bm{l}}\bm{\theta}_{\mathbb{P}}^{\bm{l}}=\sum_{P\in\mathbb{P}_{\bm{a}}}n_P\prod_{p\in P}\bm{\theta}_p.
 \end{equation*}
 
Here $n_P$ counts the total number of permutations in the sets $\mathbb{P}_i$, $i\in[m]$. This representation of non-central moments in terms of paths extends the computation of the second central moment of \citet{Sullivant2010} via the trek rule to generic non-central moments. 

\begin{example}
 Consider the expectation $\E(Y_2^2Y_4^2)$ for the variables in the DAG of Figure \ref{fig:BN}. This expectation, being the associated monomial of degree 4, can be computed by looking at all distinct tuples of rooted paths of dimension four, where two paths end in $Y_2$ and two  in $Y_4$. These are summarized in Table \ref{table:tuples}. The associated conditional expectation can be written as the following polynomial, where the $i$-th monomial corresponds to the tuple in the $i$-th row of Table $\ref{table:tuples}$:
\begin{multline*}
 \bar{u}(d\;|\;\bm{\theta})=\theta'^2_{02}\theta'^2_{04}+
 2\theta_{12}{\theta_{02}'}\theta'^2_{04}+
 \theta_{12}^2\theta'^2_{04}+
 2\theta'^2_{02}\theta_{14}\theta_{04}'+ 4\theta_{12}\theta_{02}'\theta_{14}\theta_{04}\\ 
+ 2\theta_{12}^2\theta_{14}\theta_{04}'+
 \theta'^2_{02}\theta_{14}^2+
 2\theta_{12}\theta_{02}'\theta_{14}^2+
 \theta_{12}^2\theta_{14}^2.
 \end{multline*}
 Note for example that $\theta_{12}{\theta_{02}'}\theta'^2_{04}$ has coefficient 2 since the paths $(Y_2)$ and $(Y_1,(Y_1,Y_2))$ can be permuted, whilst $\theta_{12}\theta_{02}'\theta_{14}\theta_{04}$ has coefficient 4 since both pairs of paths $(Y_2)$ and $(Y_1,(Y_1,Y_2))$ and $(Y_4)$ and $(Y_1,(Y_1,Y_4))$ can be permuted.
 \end{example}

 \begin{table}
 \renewcommand{\arraystretch}{1.5}
 \begin{center}
 \begin{tabular}{|c|}
 \hline
 $((2),(2),(4),(4))$\\
 $((1,(1,2)),(2),(4),(4))$\\
 $((1,(1,2)),(1,(1,2)),(4),(4))$\\
 $((2),(2),(1,(1,4)),(4))$\\
 $((1,(1,2)),(2),(1,(1,4)),(4))$\\
  $((1,(1,2)),(1,(1,2)),(1,(1,4)),(4))$\\
  $((2),(2),(1,(1,4)),(1,(1,4)))$\\
 $((1,(1,2)),(2),(1,(1,4)),(1,(1,4)))$\\
 $((1,(1,2)),(1,(1,2)),(1,(1,4)),(1,(1,4)))$\\
 \hline
 \end{tabular}
 \end{center}
 \caption{Tuples of dimension 4 with two paths ending in $Y_2$ and two more ending in $Y_4$ in the graph in Figure \ref{fig:BN}. \label{table:tuples}}
 \end{table}

Just as in the additive case, we are now able to deduce the independences required for score separability of an IDSS whose structural consensus includes a BN. We let $\bm{\theta}_{\Gr}^{\bm{b}}$ be the simplified version of $\bm{\theta}_{\mathbb{P}}^{\bm{a}}$ where parameters only appear once and the exponent are appropriately summed.  

\begin{theorem}
\label{theo:3}
Suppose that the CK-class of an IDSS includes a linear SEM over a DAG $\Gr$, where panel $G_i$ oversees $Y_i$, $i\in[m]$, and a panel separable  utility. Suppose panel $G_i$ agreed to use a polynomial utility function of degree $n_i$, $i\in[m]$. If, for every $\bm{b}\simeq \bm{n}$,  where $\bm{n}=(n_i)_{i\in[m]}\in\mathbb{Z}^{m}_{\geq 0}$, $\bm{\theta}_{\Gr}$ entertains moment independence of order $\bm{b}$, then the IDSS is score separable.  
\end{theorem}

The proof of this result is given in Appendix \ref{app:G}.
Theorem \ref{theo:3} ensures adequacy for a large class of IDSSs based on flexible multilinear utility factorizations and commonly used BNs defined as linear SEMs.

\section{An application in household food security}
\label{sec6}
To illustrate the application of our results in a real-world example, we consider the food security network discussed in Section \ref{sec:example} and reported in Figure \ref{fig:BN}. Suppose the variables are each under the jurisdiction of a different panel of experts. Jointly they reach a consensus to:
\begin{itemize}
\item investigate the effectiveness of an increase ($d_0$), decrease ($d_1$) or not a change ($d_2$) of the number of pupils eligible for free school meals, with $\mathbb{D}=\{d_0,d_1,d_2\}$ (decision consensus);
\item  model the conditional dependences between the four random variables deemed relevant with the BN reported in Figure \ref{fig:BN} (structural consensus);
\item consider two utility classes of utility factorizations - the first with  preferentially independent attributes (class $\mathbb{U}_1)$ as in equation (\ref{eq:last}), the second enjoying a multilinear utility factorization (class $\mathbb{U}_2)$ defined as
\[
u(d,\bm{y})=\sum_{I\in \mathcal{P}_0([4])}k_I(d)\prod_{i\in I}u_i(y_i,d).
\]
\end{itemize}
These agreements give the CK-class for this application. Next suppose that each panel decides to model the variable under its jurisdiction via a linear SEM as specified in equation (\ref{eq:sem}) and to model its marginal utility with a polynomial utility function of degree two, i.e. $u_i(y_i,d)=\rho_{i1}(d)y_i+\rho_{i2}(d)y_i^2$, $i\in[4]$.

The two questions that we next  address are the following: what independences do the panels need to be prepared to make for the IDSS to be adequate? What summaries do they have to deliver? The answer depends on the class of utility functions chosen. We thus first focus on the simpler class $\mathbb{U}_1$ of preferentially independent attributes. 

\begin{table}
\begin{center}
\begin{tabular}{llll}
\hline
$k_1\rho_{11}\theta_{01}$, & $k_1\rho_{12}\theta_{01}^2$& $k_1\rho_{12}\psi_1$& $k_2\rho_{21}\theta_{02}$\\
$k_2\rho_{21}\theta_{01}\theta_{12}$& $k_2\rho_{22}\theta_{02}^2$&$k_2\rho_{22}\psi_2$&$k_2\rho_{22}\theta_{01}^2\theta_{12}^2$\\
$k_2\rho_{22}\psi_1\theta_{12}^2$&$2k_2\rho_{22}\theta_{01}\theta_{02}\theta_{12}$& $k_3\rho_{31}\theta_{03}$&$k_3\rho_{31}\theta_{02}\theta_{23}$\\
$k_3\rho_{31}\theta_{01}\theta_{12}\theta_{23}$&$k_3\rho_{31}\theta_{01}\theta_{13}$&$k_3\rho_{32}\theta_{03}^2$&$k_3\rho_{32}\psi_3$\\
$k_3\rho_{32}\theta_{01}^2\theta_{13}^2$&$k_3\rho_{32}\theta_{13}^2\psi_1$&$k_3\rho_{32}\theta_{02}^2\theta_{23}^2$&$2k_3\rho_{32}\theta_{03}\theta_{01}\theta_{12}\theta_{23}$\\
$k_3\rho_{32}\psi_2\theta_{23}^2$&$k_3\rho_{32}\psi_1\theta_{12}^2\theta_{23}^2$&$k_3\rho_{32}\theta_{01}^2\theta_{12}^2\theta_{23}^2$&$2k_3\rho_{32}\theta_{03}\theta_{02}\theta_{23}$\\
$2k_3\rho_{32}\theta_{03}\theta_{01}\theta_{13}$&$2k_3\rho_{32}\psi_1\theta_{12}\theta_{13}\theta_{23}$&$2k_3\rho_{32}\theta_{01}\theta_{02}\theta_{13}\theta_{23}$&$2k_3\rho_{32}\theta_{12}\theta_{13}\theta_{23}\theta_{01}^2$\\
$2k_3\rho_{32}\theta_{01}\theta_{02}\theta_{12}\theta_{23}^2$&$k_4\rho_{41}\theta_{04}$&$k_4\rho_{41}\theta_{01}\theta_{14}$&$k_4\rho_{42}\theta_{04}^2$\\
$k_4\rho_{42}\psi_4$&$k_4\rho_{42}\theta_{01}^2\theta_{14}^2$&$k_4\rho_{42}\psi_1\theta_{14}^2$&$2k_4\rho_{42}\theta_{01}\theta_{04}\theta_{14}$\\
\hline
\end{tabular}
\end{center}
\caption{Monomials of the CEU for the utility class $\mathbb{U}_1$.\label{table:addeu}}
\end{table}

Through the process of algebraic substitutions, as formalized in Lemma \ref{lemma:addeu}, the CEU for the utility class $\mathbb{U}_1$ can be computed as the sum of the monomials reported in Table \ref{table:addeu} where we left the dependence on the decision $d\in\mathbb{D}$ implicit. Given this list of monomials, it is then straightforward the identify the independences required by the IDSS to be able to compute the EU of any available policy as a function of beliefs individually delivered by panels. Specifically, the  moment independences summarized in Table \ref{table:ind} need to hold. Assuming these, then each panel can deliver independently the required beliefs to derive appropriate EU scores uniquely. So, for instance, if the panels delivered the beliefs reported in Appendix \ref{app}, then the IDSS would recommend that the number of pupils eligible for free school meals is increased since the EU of this policy equals $1.87$, whilst for $d_1$ and $d_2$ this is $0.51$ and $0.62$ respectively. 

\begin{table}
\begin{center}
\begin{tabular}{ll}
\hline
$\mathbb{E}(\theta_{01}\theta_{12})=\mathbb{E}(\theta_{01})\mathbb{E}(\theta_{12})$ & $\mathbb{E}(\theta_{01}^2\theta_{12}^2)=\mathbb{E}(\theta_{01}^2)\mathbb{E}(\theta_{12}^2)$ \\
 $\mathbb{E}(\psi_1\theta_{12}^2)=\mathbb{E}(\psi_1)\mathbb{E}(\theta_{12}^2)$ & $\mathbb{E}(\theta_{01}\theta_{02}\theta_{12})=\mathbb{E}(\theta_{01})\mathbb{E}(\theta_{02}\theta_{12})$ \\
   $\mathbb{E}(\theta_{02}\theta_{23})=\mathbb{E}(\theta_{02})\mathbb{E}(\theta_{23})$ & $\mathbb{E}(\theta_{01}\theta_{12}\theta_{23})=\mathbb{E}(\theta_{01})\mathbb{E}(\theta_{12})\mathbb{E}(\theta_{23})$\\
   $\mathbb{E}(\theta_{01}\theta_{13})=\mathbb{E}(\theta_{01})\mathbb{E}(\theta_{13})$ & $\mathbb{E}(\theta_{01}^2\theta_{13}^2)=\mathbb{E}(\theta_{01}^2)\mathbb{E}(\theta_{13}^2)$\\
  $\mathbb{E}(\theta_{13}^2\psi_1)=\mathbb{E}(\theta_{13}^2)\mathbb{E}(\psi_1)$&  $\mathbb{E}(\theta_{02}^2\theta_{23}^2)=\mathbb{E}(\theta_{02}^2)\mathbb{E}(\theta_{23}^2)$ \\
  $\mathbb{E}(\theta_{01}\theta_{12}\theta_{03}\theta_{23})=\mathbb{E}(\theta_{01})\mathbb{E}(\theta_{12})\mathbb{E}(\theta_{03}\theta_{23}) $&   $\mathbb{E}(\theta_{23}^2\psi_2)=\mathbb{E}(\theta_{23}^2)\mathbb{E}(\psi_2)$\\
  $\mathbb{E}(\psi_1\theta_{12}^2\theta_{23}^2)=\mathbb{E}(\psi_1)\mathbb{E}(\theta_{12}^2)\mathbb{E}(\theta_{23}^2)$ &   $\mathbb{E}(\theta_{01}^2\theta_{12}^2\theta_{23}^2)=\mathbb{E}(\theta_{01}^2)\mathbb{E}(\theta_{12}^2)\mathbb{E}(\theta_{23}^2)$\\
 $\mathbb{E}(\theta_{02}\theta_{03}\theta_{23})=\mathbb{E}(\theta_{02})\mathbb{E}(\theta_{03}\theta_{23})$ &   $\mathbb{E}(\theta_{01}\theta_{03}\theta_{13})=\mathbb{E}(\theta_{01})\mathbb{E}(\theta_{03}\theta_{13})$ \\
$ \mathbb{E}(\psi_1\theta_{12}\theta_{13}\theta_{23})= \mathbb{E}(\psi_1)\mathbb{E}(\theta_{12})\mathbb{E}(\theta_{13}\theta_{23})$ & $ \mathbb{E}(\theta_{01}\theta_{02}\theta_{13}\theta_{23})= \mathbb{E}(\theta_{01})\mathbb{E}(\theta_{02})\mathbb{E}(\theta_{13}\theta_{23})$ \\
$ \mathbb{E}(\theta_{01}\theta_{02}\theta_{12}\theta_{23}^2)= \mathbb{E}(\theta_{01})\mathbb{E}(\theta_{02}\theta_{12})\mathbb{E}(\theta_{23}^2)$ & $ \mathbb{E}(\theta_{12}\theta_{13}\theta_{23}\theta_{01}^2)= \mathbb{E}(\theta_{12})\mathbb{E}(\theta_{13}\theta_{23})\mathbb{E}(\theta_{01}^2)$ \\
$\mathbb{E}(\theta_{01}\theta_{14})=\mathbb{E}(\theta_{01})\mathbb{E}(\theta_{14})$ & $\mathbb{E}(\theta_{01}^2\theta_{14}^2)=\mathbb{E}(\theta_{01}^2)\mathbb{E}(\theta_{14}^2)$\\
$\mathbb{E}(\psi_1\theta_{14}^2)=\mathbb{E}(\psi_1)\mathbb{E}(\theta_{14}^2)$ & $\mathbb{E}(\theta_{01}\theta_{04}\theta_{14})=\mathbb{E}(\theta_{01})\mathbb{E}(\theta_{04}\theta_{14})$ \\
\hline
\end{tabular}
\end{center}
\caption{Moment independences required by the IDSS for adequacy. \label{table:ind}}
\end{table}

We next consider the case where the CK-class includes the second class $\mathbb{U}_2$ of multilinear utilities. Whilst for preferentially independent attributes the CEU has 36 monomials (in Table \ref{table:addeu}), in this case the CEU can be shown to have 3869 monomials. In this case the computer algebra software Mathematica instantaneously gives us the CEU polynomial using the simple code reported in  Appendix \ref{app:code}. The output CEU can then be used to identify the required moment independences and panels' beliefs. For instance, the parameter $\theta_{01}$ has degree up to 8 in the polynomial CEU when $\mathbb{U}_2$ is used, whilst for $\mathbb{U}_1$ its maximum degree was 2. Using this more general class of utilities and the specifications given in Appendix \ref{appp}, the IDSS would again recommend that the number of pupils eligible for free school meals is increased since the EU of this policy equals $0.97$, whilst for $d_1$ and $d_2$ this is $0.16$ and $0.37$ respectively. 

\section{Discussion}
The framework of IDSSs is capable of supporting decision making in situations where judgements come from different panels of experts having jurisdiction over different aspects of the system. In this paper we have relaxed many of the assumptions guaranteeing coherence in this type of systems \citep{Smith2015} by exploiting the polynomial structure of certain statistical models and utility functions and illustrated their usefulness in a practical application.

In the particular context where the structural consensus includes a BN model, the process of algebraic substitution has proven fundamental in identifying the required summaries and independence relations. We have encouraging results, to be reported in future work, towards a generalization of such recursions in dynamic models, as the multiregression dynamic model \citep{Queen1993}, where expressions for the moments' forecasts can be deduced in closed form. Furthermore, when each vertex of the BN is no longer a random variable but a random vector (for example when a variable is measured at different geographic location), the theory of tensors \citep{McCullagh1987} can be employed to concisely report the associated EU expressions. We plan to develop such a methodology in future work.

%\begin{acknowledgements}
%\end{acknowledgements}

\bibliographystyle{spbasic}      % basic style, author-year citations
\bibliography{bib1.bib}   % name your BibTeX data base

\appendix
\section{Proofs}
\subsection{Proof of Corollary \ref{lemma:ciao3}}
\label{app:A}
Adequacy is guaranteed if the EU function can be written in terms of $\mu_{ji}(d)$ and $k_{\bm{b}}(d)$, $i\in[m]$, $j\in[s_i]$ and $d\in\mathcal{D}$. Note that
\begin{eqnarray*}
\bar{u}(d)&=&\E\left(q_{d}(\bm{\lambda }_{_{1}}(\bm{\theta }_{1},d),\dots,\bm{\lambda }_{m}(\bm{\theta }_{m},d))\right)\\
&=&\sum_{\bm{b}\in B}k_{\bm{b}}(d)\E\Big( \prod_{i\in[m]}\prod_{j\in[s_i]^0}\lambda_{ji}(\bm{\theta }_{i},d)^{b_{j,i}}\Big)\\
&=&\sum_{\bm{b}\in B}k_{\bm{b}}(d)\E\Big(\prod_{i\in[m]}\prod_{j\in[s_i]^0}\bm{\theta}_i^{\bm{a}_{ji}}\Big).
\end{eqnarray*}
The argument of this expectation is a monomial of multi-degree lower or equal to $\bm{a}^*$. Moment independence then implies that 
$
\bar{u}(d)=\sum_{\bm{b}\in B}k_{\bm{b}}(d)\prod_{i\in[m]}\prod_{j\in[s_i]^0}\mu_{ji}(d),
$ 
and the result follows. 
\subsection{Proof of Theorem \ref{theo:ciao}}
\label{app:B}
Fix a policy $d\in\mathbb{D}$ and suppress this dependence. Under the assumptions of the theorem, the utility function can be written as
\begin{equation}
u(\bm{y})=\sum_{I\in\mathcal{P}_0([m])}k_I\sum_{i\in I}\left(\sum_{j\in [n_i]}\rho_{ij}y_i^{j}\right).
\label{eq:prova}
\end{equation}
Note also that we can rewrite (\ref{eq:prova}) as
\begin{equation*}
u(\bm{y})=\hat{u}(\bm{y}_{[m-1]})+\hat{u}(y_{m}),
\end{equation*}
where 
\begin{align}
\hat{u}(\bm{y}_{[m-1]})&=\sum_{I\in\mathcal{P}_0([m-1])}k_I\hspace{0.2cm}\prod_{i\in I}\Big(\sum_{j\in [n_i]}\rho_{ij}y_i^{j}\Big),\nonumber\\
\hat{u}(y_{m})&=\sum_{I\in\mathcal{P}^m_0([m])}k_I\hspace{0.1cm}\prod_{i\in I}\Big(\sum_{j\in [n_i]}\rho_{ij}y_i^{j}\Big),\label{eq:prova1}
\end{align}
and $\mathcal{P}^m_0([m])=\mathcal{P}_0([m])\cap\{m\}$.
Calling $\bm{\theta}$ the overall parameter vector of the IDSS, the CEU function can be written applying sequentially the tower rule of expectation as
\begin{equation}
\E(u(\bm{Y})\;|\;\bm{\theta})=\E_{Y_1|\bm{\theta}}\Big(\cdots \E_{Y_{m{-1}}|\bm{Y}_{[m{-2}]},\bm{\theta}}\big(\hat{u}(\bm{y}_{[m{-1}]})+\E_{Y_m|\bm{Y}_{[m{-1}]},\bm{\theta}}(\hat{u}(\bm{y}_{m}))\big)\Big).
\label{eq:prova3}
\end{equation}
From equation (\ref{eq:prova1}), the definition of a polynomial SEM and observing that the power of a polynomial is still a polynomial function in the same arguments, it follows that 
\[
E_{Y_m|\bm{Y}_{[m-1]},\bm{\theta}}\left(\hat{u}(\bm{y}_{m})\right)=p_m(\bm{Y}_{[m-1]},\bm{\theta}), 
\]
where $p_m$ is a generic polynomial function. Thus $\hat{u}(\bm{Y}_{[m-1]})+\E_{Y_m|\bm{Y}_{[m-1],\bm{\theta}}}\left(\hat{u}(\bm{y}_{m})\right)$ is also a polynomial function in the same arguments. 
Following the same reasoning, we have that 
\begin{equation*}
\E_{Y_{m-1}|\bm{Y}_{[m-2]},\bm{\theta}}\left(\hat{u}(\bm{y}_{[m-1]})+\E_{Y_m|\bm{Y}_{[m-1]},\bm{\theta}}\left(\hat{u}(\bm{y}_{m})\right)\right)=p_{m-1}(\bm{Y}_{[m-2]},\bm{\theta}),
\end{equation*}
where $p_{m-1}$ is a generic polynomial function. Therefore the same procedure can be applied to all the expectations in (\ref{eq:prova3}). So $
\E(u(\bm{Y})\;|\;\bm{\theta})=p_1(\bm{\theta}),$  
 where $p_1$ is a generic polynomial function. This defines by construction an algebraic CEU, where the functions $\lambda_{ij}$ are monomials. Quasi independence and Lemma \ref{lemma:ciaociao} then guarantee score separability holds. 

\subsection{Proof of Proposition \ref{prop:algsub}}
\label{app:C}
We prove equation (\ref{eq:1}) via induction over the indices of the variables. Let $Y_1$ be a root of $\Gr$. Thus $Y_1=\theta_{01}'$, where $\theta_{01}'$ is the monomial associated to the only rooted path ending in $Y_1$, namely $(Y_1)$. Assume the result is true for $Y_{n-1}$ and consider $Y_{n}$. By the inductive hypothesis we have that, if $i<j$ whenever $i\in \Pi_j$, 
\begin{equation}
Y_{n}=\theta_{0n}'+\sum_{i\in \Pi_n}\theta_{in}Y_i=\theta_{0n}'+\sum_{i\in \Pi_n}\theta_{in}\sum_{P\in \mathbb{P}_i}\bm{\theta}_P.
\label{eq:pathid}
\end{equation}
Note that every rooted path ending in $Y_n$ is either $(Y_n)$ or consists of a rooted path ending in $Y_i$, $i\in \Pi_n$, together with the edge $(Y_i,Y_n)$. From this observation the result then follows by rearranging the terms in equation (\ref{eq:pathid}). Equation (\ref{eq:2})  can be proven via the same inductive process  noting that $\E(Y_1\;|\;\bm{\theta},d)=\theta_{01}'$ and  $\E(Y_n\;|\;\bm{\theta},d)=\theta_{0n}'+\sum_{i\in\Pi_n}\theta_{in}\E(Y_i\;|\;\bm{\theta},d)$.

\subsection{Proof of Theorem \ref{theo:pr}}
\label{app:E}
Under the assumptions of the theorem, the CEU function can be written as in equation (\ref{eq:addeu}). From the linearity of the expectation operator we have that
\begin{eqnarray*}
\E(\bar{u}(d\;|\;\bm{\theta}))&=& \sum_{i\in[m],j\in[n_i]}k_i(d)\rho_{ij}(d)\sum_{|\bm{a}_i|=j}\binom{j}{\bm{a}_i}\E\left(\bm{\theta}_{\mathbb{P}_i}^{\bm{a}_i}\right)\\
&=&\sum_{i\in[m],j\in[n_i]}k_i(d)\rho_{ij}(d)\sum_{|\bm{c}_i|=j}\binom{j}{\bm{c}_i}\E\left(\bm{\theta}_{\Gr_i}^{\bm{c}_i}\right).
\end{eqnarray*}
Applying moment independence and letting $V_i$ and $E_i$ be the sets of distinct vertices and edges, respectively, for all the elements $P\in\mathbb{P}_i$, we have that  
\[
\E(\bar{u}(d\;|\;\bm{\theta}))= \hspace{-.1cm}\sum_{\substack{i\in[m], j\in[n_i],\\ |\bm{c}_i|=j}}\hspace{-.1cm}k_i(d)\rho_{ij}(d)\binom{j}{\bm{c}_i}\prod_{l\in V_i}\hspace{-.1cm}\E\left(\theta_{0l}'^{c_{il}}\theta_{lCh_l}^{c_{iCh_l}}\right)\hspace{-.1cm}\prod_{(j,k)\in E_i\setminus (l,Ch_l)}\hspace{-.1cm}\E\left(\theta_{jk}^{c_{ik}}\right),
\]
where $c_{ik}$ is the element of $\bm{c}_i$ associated to $\theta_{jk}$ and $Ch_l$ is the index of a children of the vertex $l$. The thesis  then follows since each of these expectations is delivered by an individual panel.

\subsection{Proof of Lemma \ref{lemma:multieu}}
\label{app:F}
To prove this result we first show that under the assumptions of the lemma the utility function can be written as
\begin{equation}
\label{uno}
u(\bm{y},d)=\sum_{\bm{0}<_{lex}\bm{a}\leq_{lex}\bm{n}}c_{\bm{a}}(d)\bm{y}^{\bm{a}},
\end{equation}
and then prove that 
\begin{equation}
\label{due}
\bm{Y}^{\bm{a}}=\sum_{\bm{l}\simeq \bm{a}}\binom{|\bm{a}|}{\bm{l}}\bm{\theta}_{\mathbb{P}}^{\bm{l}}.
\end{equation}
The lemma then  follows by substituting into equation (\ref{uno}) for $\bm{y}^{\bm{a}}$ given in equation (\ref{due}).

Fix a policy $d\in\mathbb{D}$ and suppress this dependence. We prove equation (\ref{uno}) via induction over the number of vertices of the DAG. If the DAG has only one vertex then
\begin{equation*}
u(\bm{y})=k_1\sum_{i\in n_1}\rho_{1i}y_1^i.
\end{equation*}
This can be seen as an instance of equation (\ref{uno}). Assume the result holds for a network with $n-1$ vertices. A multilinear utility factorisation can be rewritten as
\begin{equation}
u(\bm{y})=\sum_{I\in\mathcal{P}_0([n-1])}k_I\prod_{i\in I}u_i(y_i)+\sum_{I\in\mathcal{P}^n_0([n])}k_I\prod_{i\in I\setminus\{n\}}u_i(y_i)u_n(y_n)+ k_nu_n(y_n).
\label{macheneso}
\end{equation}
The first term on the rhs of (\ref{macheneso}) is by inductive hypothesis equal to the sum of all the possible monomial of degree $\bm{a}=(a_1,\dots,a_{n-1},0)$ where $0\leq a_i\leq n_i$, $i\in[n]$. The other terms only include monomials such that the exponent of $y_n$ is not zero. Letting $\bm{n}_{n-1}=(n_i)_{i\in[n-1]}$, $\bm{y}_{[n-1]}=\prod_{i\in[n-1]}y_i$ and $u'=\sum_{I\in\mathcal{P}_0^n([n])}k_I\prod_{i\in I\setminus \{n\}}u_i(y_i)u_n(y_n)+k_nu_n(y_n)$, we now have that
\begin{eqnarray}
u'&=& \sum_{\bm{0}<_{lex}\bm{a}\leq_{lex}\bm{n}_{n-1}}c_{\bm{a}}\bm{y}^{\bm{a}}_{[n-1]}\bigg(\sum_{i\in [n_n]}\rho_{ni}y_n^i\bigg)+k_nu_n(y_n)\nonumber\\
&=&\sum_{\substack{\bm{0}<_{lex}\bm{a}\leq_{lex}\bm{n}_{n-1}\\ i\in[n_n]}}c_{\bm{a}}\rho_{ni}\bm{y}^{\bm{a}}_{[n-1]}y_n^i+k_nu_n(y_n)=\sum_{\substack{\bm{0}'<_{lex}\bm{a}\leq_{lex}\bm{n}_{n}\\ a_n\neq 0}}c_{\bm{a}}\bm{y}_{[n]}^{\bm{a}}.\label{eq:uff}
\end{eqnarray}
Therefore, equation (\ref{uno}) follows from equations (\ref{macheneso}) and (\ref{eq:uff}). To prove equation (\ref{due}) note that the monomial $\bm{Y}^{\bm{a}}$ can be written as
\begin{align*}
\bm{Y}^{\bm{\alpha}}&=\prod_{i\in[m]}Y_i^{a_i}=\prod_{i\in[m]}\left(\sum_{|\bm{l}_i|=a_i}\binom{a_i}{\bm{l}_i}\bm{\theta}_{\mathbb{P}_i}^{\bm{l}_i}\right)=\sum_{\bm{l}\simeq \bm{a}}\bm{\theta}_{\mathbb{P}}^{\bm{l}}\prod_{i\in[m]}\binom{a_i}{\bm{l}_i}.
\end{align*}
Equation (\ref{due}) then follows by noting that 
\begin{equation*}
\prod_{i\in [m]}\binom{a_i}{\bm{l}_i}=\frac{\prod_{i\in[m]}a_i!}{\prod_{i\in[m]}\prod_{j\in[n_i]}l_{ij}!}=\binom{|\bm{a}|}{\bm{l}}.
\end{equation*}

\subsection{Proof of Theorem \ref{theo:3}}
\label{app:G}
Under the conditions of the theorem, the CEU function can be written as in (\ref{eq:multieu}). The linearity of the expectation operator than implies that
\[
\E(\bar{u}(d\;|\;\bm{\theta}))=\sum_{\substack{\bm{0}<_{lex}\bm{a}\leq_{lex}\bm{n}\\\bm{l}\simeq \bm{a}}}c_{\bm{a}}\binom{|\bm{a}|}{\bm{l}}\E\left(\bm{\theta}_{\mathbb{P}}^{\bm{l}}\right)=\sum_{\substack{\bm{0}<_{lex}\bm{b}\leq_{lex}\bm{n}\\\bm{l}\simeq \bm{b}}}c_{\bm{b}}\binom{|\bm{b}|}{\bm{l}}\E\left(\bm{\theta}_{{\Gr}}^{\bm{l}}\right).
\]
Applying moment independence and letting  $V_{tot}$ and $E_{tot}$ be the sets of distinct vertices and edges, respectively, for all the elements $P\in\mathbb{P}=\cup_{i\in[m]}\mathbb{P}_i$, we then have that for any $\bm{l}\simeq \bm{b}$
\begin{eqnarray*}
\E\left(\bm{\theta}_{\Gr}^{\bm{l}}\right)= \prod_{t\in V_{tot}}\E\left(\theta_{0t}'^{l_{it}}\theta_{tCh_t}^{l_{iCh_t}}\right)\hspace{-.1cm}\prod_{(j,k)\in E_{tot}\setminus (t,Ch_t)}\hspace{-.1cm}\E\left(\theta_{jk}^{l_{ik}}\right).
\end{eqnarray*}
 Score separability then follows since each of these expectations is delivered by an individual panel.

\section{Numerical specifications for the food security example}
\subsection{Utility class $\mathbb{U}_1$}
\begin{itemize}
\item Probabilistic panel specifications that depend on the decision taken:
\begin{center}
\scalebox{0.8}{
\begin{tabular}{c|c|c|c|c|c|c|c}
&$\E(\theta_{01})$& $\E(\psi_1)$&$\E(\theta_{04})$&$\E(\psi_4)$&$\E(\theta_{02})$& $\E(\psi_2)$&$\E(\theta_{12})$\\
\hline
$d_0$&1.5&5&30&8&5&40&7\Tstrut\Bstrut\\
$d_1$&-2&4&-5&5&-6&20&2\Tstrut\Bstrut\\
$d_2$&-0.5&3&10&4&3&15&7\Tstrut\Bstrut
\end{tabular}}
\end{center}
\item Probabilistic panel specifications independent of the decision taken:
\begin{center}
\scalebox{0.8}{
\begin{tabular}{|ccccc|}
\hline
$\E(\theta_{03})=5$, &$\E(\theta_{13})=17$,&$\E(\theta_{23})=10$,&$\E(\theta_{14})=10$,&$\E(\psi_{3})=20$,\Tstrut\Bstrut\\
$\V(\theta_{01})=1$,&$\V(\theta_{02})=1$,&$\V(\theta_{03})=1$,&$\V(\theta_{04})=1$,&$\V(\theta_{12})=1$,\Tstrut\Bstrut\\
$\V(\theta_{12})=1$,&$\V(\theta_{13})=3$,&$\V(\theta_{14})=2$,&$\V(\theta_{23})=2$,&\Tstrut\Bstrut\\
\hline
\end{tabular}}
\end{center}
\item Criterion weights and terms in the utility functions\footnote{Notice that these values are then normalized to give utility functions between 0 and 1.}
\begin{center}
\scalebox{0.8}{
\begin{tabular}{|cccc|}
\hline
$k_1=0.25$, &$k_2=0.25$,&$k_3=0.25$, &$k_4=0.25$, \Tstrut\Bstrut\\
$\rho_{11}=-2$, &$\rho_{12}=1$,&$\rho_{21}=2$, &$\rho_{22}=10$, \Tstrut\Bstrut\\
$\rho_{31}=8$, &$\rho_{32}=0.5$,&$\rho_{41}=3$, &$\rho_{42}=-5$, \Tstrut\Bstrut\\
\hline
\end{tabular}}
\end{center}
\end{itemize}

\label{app}
\subsection{Utility class $\mathbb{U}_2$}
\label{appp}
In the multilinear case higher moments are required. Here we assume that these can be computed from the first two moments in Appendix \ref{app} using the recursions of normal distributions. The only specifications that change for this second class are the criterion weights given in the following table.

\begin{center}
\scalebox{0.8}{
\begin{tabular}{|ccccc|}
\hline
$k_1=0.15$, &$k_2=0.15$,&$k_3=0.15$, &$k_4=0.15$, &$k_{12}=0.05$,\Tstrut\Bstrut\\
$k_{13}=0.05$,&$k_{14}=0.05$, &$k_{23}=0.05$, &$k_{24}=0.05$, &$k_{34}=0.05$,\Tstrut\Bstrut\\
$k_{123}=0.02$, &$k_{124}=0.02$, &$k_{134}=0.02$,&$k_{234}=0.02$,  &$k_{1234}=0.02$, \Tstrut\Bstrut\\
\hline
\end{tabular}}
\end{center}

\section{Code for the multilinear factorization}
\label{app:code}
\texttt{y4 := t04 + t14*y1 + e4;}\\
\texttt{y3 := t03 + t13*y1 + t23*y2 + e3;}\\
\texttt{y2 := t02 + t12*y1 + e2;}\\
\texttt{y1 := t01 + e1;}\\
\texttt{u4 := c4*y4$\wedge$2 + b4*y4;}\\
\texttt{u3 := c3*y3$\wedge$2 + b3*y3;}\\
\texttt{u2 := c2*y2$\wedge$2 + b2*y2;}\\
\texttt{u1 := c1*y1$\wedge$2 + b1*y1;}\\
\texttt{u := k1*u1 + k2*u2 + k3*u3 + k4*u4 + k12*u1*u2 + k13*u1*u3 + 
   k14*u1*u4}\\ 
\texttt{+ k23*u2*u3 + k24*u2*u4 + k34*u3*u4 + k123*u1*u2*u3 + 
  k124*u1*u2*u4}\\
\texttt{+ k134*u1*u3*u4 + k234*u2*u3*u4 + k1234*u1*u2*u3*u4;}\\
\texttt{eu4 := ReplaceAll[Collect[u, e4], \{e4 -> 0, e4$\wedge$2 -> psi4\}];}\\
\texttt{eu3 := ReplaceAll[Collect[eu4, e3], \{e3 -> 0, e3$\wedge$2 -> psi3\}];}\\
\texttt{eu2 := ReplaceAll[Collect[eu3, e2], \{e2 -> 0, e2$\wedge$2 -> psi2\}];}\\
\texttt{eu1 := ReplaceAll[Collect[eu2, e1], \{e1 -> 0, e1$\wedge$2 -> psi1\}];}\\
\texttt{eu1}\\

%\begin{thebibliography}{}
% etc
%\end{thebibliography}

\end{document}